\documentclass[useAMS,usenatbib,usegraphicx,onecolumn]{mn2e}
\usepackage{graphicx}                                            
\usepackage{times}
\usepackage{float}
\usepackage{rotating}
\usepackage{epstopdf}
\usepackage{multirow}
\usepackage{pdflscape}
\usepackage{color}
\usepackage{tabularx}

\def\ltsima{$\; \buildrel < \over \sim \;$}
\def\simlt{\lower.5ex\hbox{\ltsima}}
\def\gtsima{$\; \buildrel > \over \sim \;$}
\def\simgt{\lower.5ex\hbox{\gtsima}}
\def\gsimeq
{\hbox{\raise0.5ex\hbox{$>\lower1.06ex\hbox{$\kern-1.07em{\sim}$}$}}}
\def\lsimeq
{\hbox{\raise0.5ex\hbox{$<\lower1.06ex\hbox{$\kern-1.07em{\sim}$}$}}}

\def\xmm{{\it XMM-Newton }}

\def\xmm{{\it XMM-Newton}}
\def\chandra{{\it Chandra}}

\def\fermi{{Fermi}}

\def\sgras{Sgr~A$^{\star}$}

\def\xis{XIS}
\def\xis1{XIS1}
\def\xis2{XIS2}
\def\xis3{XIS3}

\title[] 
 {{Reply to comment on: "An X-ray chimney extending hundreds of parsecs above and below the Galactic Centre"}}
 \author[G.\ Ponti et al. ]
  {G.~Ponti$^{1,2}$, F.~Hofmann$^{1}$, E.~Churazov$^{3,4}$, M.~R.~Morris$^{5}$, F.~Haberl$^{1}$, K.~Nandra$^{1}$, 
     \newauthor
  R.~Terrier$^{6}$, M.~Clavel$^{7}$, and A.~Goldwurm$^{6,8}$ \\
\\  
 $^1$ Max-Planck-Institut f\"ur Extraterrestrische Physik, Giessenbachstrasse 1, D-85748 Garching, Germany \\
 $^2$ INAF-Osservatorio Astronomico di Brera, Via E. Bianchi 46, I-23807 Merate (LC), Italy \\
 $^3$ Max-Planck-Institut f\"ur Astrophysik, Karl-Schwarzschild-Str. 1, D-85748, Garching, Germany \\
 $^4$ Space Research Institute (IKI), Profsoyuznaya 84/32, Moscow 117997, Russia \\
 $^5$ Department of Physics and Astronomy, University of California, Los Angeles, CA 90095-1547, USA \\
 $^6$ Unit\'e mixte de recherche Astroparticule et Cosmologie, 10 rue Alice Domon et L\'eonie Duquet, F-75205 Paris, France \\
 $^7$ Univ. Grenoble Alpes, CNRS, IPAG, F-38000 Grenoble, France \\
 $^8$ Service d'Astrophysique (SAp), IRFU/DSM/CEA-Saclay, F-91191 Gif-sur-Yvette Cedex, France
}

\pagerange{\pageref{firstpage}--\pageref{lastpage}}
\usepackage{times}

\begin{document}
\label{firstpage}
 \maketitle
In response to the comment posted by Nakashima et al. (arXiv:1903.1176), 
regarding prior claims for the features that we referred to as the Galactic 
Center Chimneys (2019, Nature, 567, 347), we point out the following: 

\begin{enumerate}
\item{} The Nakashima et al. 2019 paper appeared in the arXiv on March 8$^{th}$
(1903.02571), after our paper was in the final stage of printing (accepted 
on January 30$^{th}$). It is however interesting to see that the morphology of the 
brightest portions of the two results are in broad agreement (compare their 
Fig. 1 to our Extended Data Figs. 1 and 2). 

\item{} Nakashima et al. 2013 ApJ 773, 20 claim the discovery of a blob 
of recombining plasma $\sim1^\circ$ south of \sgras, implying peculiar abundances. 
Again, their image (Fig. 1) agrees with the brightest portions of our images,
although it does not show any direct connection between the plasma blob 
and the central parsec (e.g., such as the quasi-continuous chimney that we
reported), nor evidence for an outflow from the center. We apologize for 
overlooking an appropriate citation to this contribution by Nakashima et al.

\item{} We fitted the \xmm\ and \chandra\ data at the same position of the claimed
recombining plasma and we did not find any clear-cut evidence for the
presence of either an over-ionised plasma or peculiar abundances. Future 
X-ray calorimetric observations will presumably clarify this disagreement. 

\item{} The continuity of the Chimney features, their quasi-symmetrical placement 
relative to \sgras, and their relatively sharp and well-defined edges are the
essential features of our data that have led us to propose that the Chimneys
are a unified columnar structure that represents a channel for the outflow
of energy from the central region, possibly contributing to the stocking of the 
relativistic particle population manifested in the \fermi\ Bubbles.

\end{enumerate}

\end{document}